\documentclass[12pt]{iopart}

\newcommand{\Xmax}{X_{\mathrm{max}}}
\usepackage{lineno} 
\usepackage{graphicx}
\begin{document}

\title[UHECRs at the Pierre Auger Observatory]{The science of ultra-high energy cosmic rays after more than 15 years of operation of the Pierre Auger Observatory}

\author{Olivier Deligny\textsuperscript{*},
for the Pierre Auger Collaboration\textsuperscript{**}}

\address{\bf \textsuperscript{*} Laboratoire de Physique des 2 Infinis Ir\`ene Joliot-Curie (IJCLab)\\
CNRS/IN2P3, Universit\'{e} Paris-Saclay, Orsay, France\\
\bf \textsuperscript{**} Full author list: https://www.auger.org/archive/authors\_2022\_09.html}
\ead{\bf \textsuperscript{*} deligny@ijclab.in2p3.fr, \bf \textsuperscript{**}  spokespersons@auger.org}
\vspace{10pt}
\begin{indented}
\item[]September 2022
\end{indented}

\begin{abstract}
The Pierre Auger Observatory has been detecting ultra-high energy cosmic rays (UHECRs) for more than fifteen years. An essential feature of the Observatory is its hybrid design: cosmic rays above $100~$PeV are detected through the observation of the associated air showers with different and complementary techniques, from surface detector arrays and fluorescence telescopes to radio antennas. The analyses of the multi-detector data have enabled high-statistics and high-precision studies of the energy spectrum, mass composition and distribution of arrival directions of UHECRs. The resulting picture is summarized in this contribution. While no discrete source of UHECRs has been identified so far, the extragalactic origin of the particles has been recently determined from the arrival directions above 8~EeV, and the ring is closing around nearby astrophysical sites at higher energies. Also, the established upper limits on fluxes of UHE neutrinos and photons have implications on dark matter and cosmological aspects that are also presented in this contribution.
\end{abstract}

%
%
%
%
%

\textit{The Pierre Auger Observatory.} Ultra-high energy cosmic rays (UHECRs) are the most energetic particles produced in nature, with energies in excess of $100~$EeV. Uncovering their origin is a persistent task since their discovery, mostly because of the very small value of their intensity on Earth and because of the magnetic deflections they experience en route to Earth. The Pierre Auger Observatory, located in the province of Mendoza (Argentina) and covering 3000 km$^2$, is the present flagship experiment studying UHECRs~\cite{PierreAuger:2015eyc}. Two techniques of detection are combined to measure the extensive air showers (EAS). A surface detector (SD) array, consisting of 1660 autonomously operated water-Cherenkov detectors, provides a lateral sampling of the EAS at the ground level. The detectors are arranged on a triangular grid of 1500~m spacing (SD-1500~m), except for a denser infill area of $\simeq30~$km$^2$, where the spacing is 750~m (SD-750~m). The atmosphere above the SD array is overlooked by fluorescence detectors (FD), which consists of telescopes that detect the faint UV light emitted by nitrogen molecules previously excited by the charged particles from the EAS. This technique provides the opportunity of performing shower calorimetry by mapping the ionization content along the shower tracks, and of measuring the primary energy on a nearly model-independent basis. The FD can only operate during dark, moonless nights with a field of view free of clouds. Online and long-term performances of the detectors and data quality are monitored continuously, and a set of high-quality devices installed in the Observatory array monitor the atmospheric conditions during operation.  
 
The SD  operates with a quasi-permanent duty cycle and thus provides a harvest of data. Yet, assessing the energy of the observed events requires assumptions about the primary mass and the hadronic processes that control the cascade development. This proves to be a difficult task as the primary mass on an event-by-event basis is unknown and the centre-of-mass energy reached at the LHC corresponds only to that of a proton of $\simeq 100~$PeV colliding with a nitrogen nucleus. To circumvent these limitations, the energies are obtained by making use of a subset of events detected simultaneously by the FD and the SD.  This ``hybrid'' approach allows a calorimetric estimate of the energy for events recorded during periods when the FD cannot be operated.  

\begin{figure}[!t]
  \centering
  \includegraphics[width=10cm]{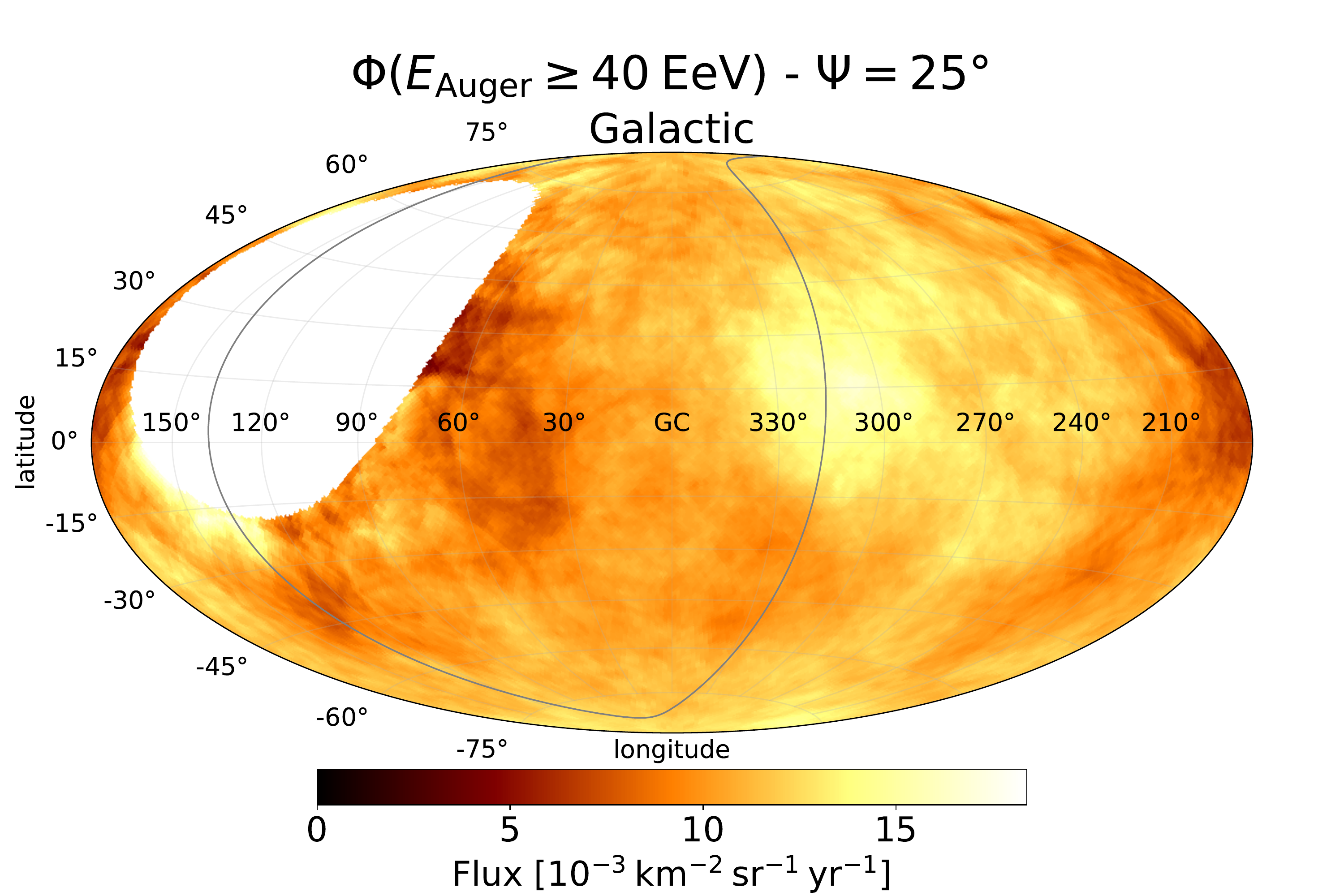}
  \caption{Flux map at energies above 40 EeV with a top-hat smoothing radius of 25$^\circ$ in Galactic coordinates. The supergalactic plane is shown as a gray line. The blank area is outside the field of view of the  Observatory. From~\cite{PierreAuger:2022axr}.}
  \label{fig:skymap}
\end{figure}

\textit{Arrival directions.} The identification of UHECR sources relies primarily on capturing in the arrival directions a pattern suggestive in an evident way of a class of astrophysical objects. Such a capture is still eluding our grasp, but some recent observations have confirmed the long-lasting broad statement of the extragalactic origin of UHECRs above the so-called ankle energy. On the one hand, an anisotropy at large scales has been revealed above ${\simeq}\:8\:$EeV~\cite{PierreAuger:2017pzq}, the amplitude and the direction of which are consistent with expectations drawn from sources distributed in a similar manner to the extragalactic matter~\cite{PierreAuger:2017pzq,PierreAuger:2018zqu}. On the other hand, at higher energies, the energy losses of UHECRs limit the horizon of the highest-energy particles. For small-enough magnetic deflections, the distribution of the arrival directions of UHECRs above $\simeq 40~$EeV could mirror the inhomogeneous distribution of the nearby extragalactic matter. Data have been subjected to comprehensive anisotropy searches for different energy thresholds above $32~$EeV, and within different angular windows, between $1^\circ$ and $30^\circ$~\cite{PierreAuger:2022axr}. Searches for significant excesses anywhere in the sky have been performed, as well as searches for correlations with known astrophysical structures and with objects that are considered plausible candidates for UHECR sources. Out of all the searches performed, above $\simeq~38~$EeV, the flux patterns of massive, star-forming of active galaxies within 200~Mpc provide a 4$\sigma$ evidence for anisotropy~\cite{PierreAuger:2022axr}. Overall, the ring is closing around nearby astrophysical sites to explain the main features of the flux sky map shown in Galactic coordinates in Fig.~\ref{fig:skymap}, obtained by filtering the data at $25^\circ$ above 40~EeV~\cite{PierreAuger:2022axr}.

\textit{Energy spectrum and mass composition.} The energy spectrum and chemical composition of UHECRs observed on Earth result from the emission processes at play, which encompass acceleration mechanisms, losses and escape from the source environments as well as propagation effects. Differently from, and complementary to anisotropies, these two observables provide constraints helping infer the properties of the acceleration processes, the energetics of the sources, and the abundances of elements in the source environments.

\begin{figure}[!t]
  \centering
  \includegraphics[width=0.5\textwidth]{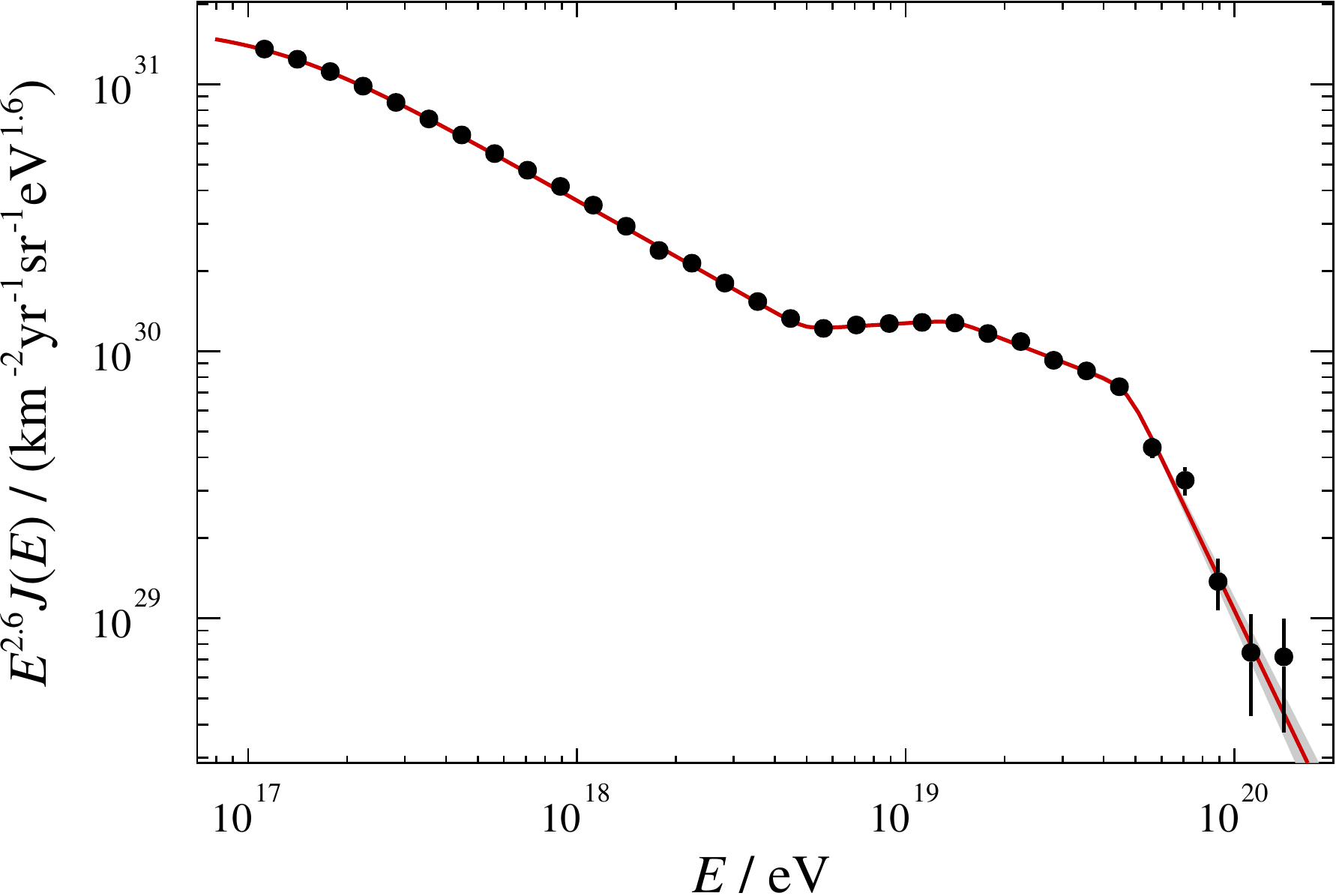}
  \includegraphics[width=0.6\textwidth]{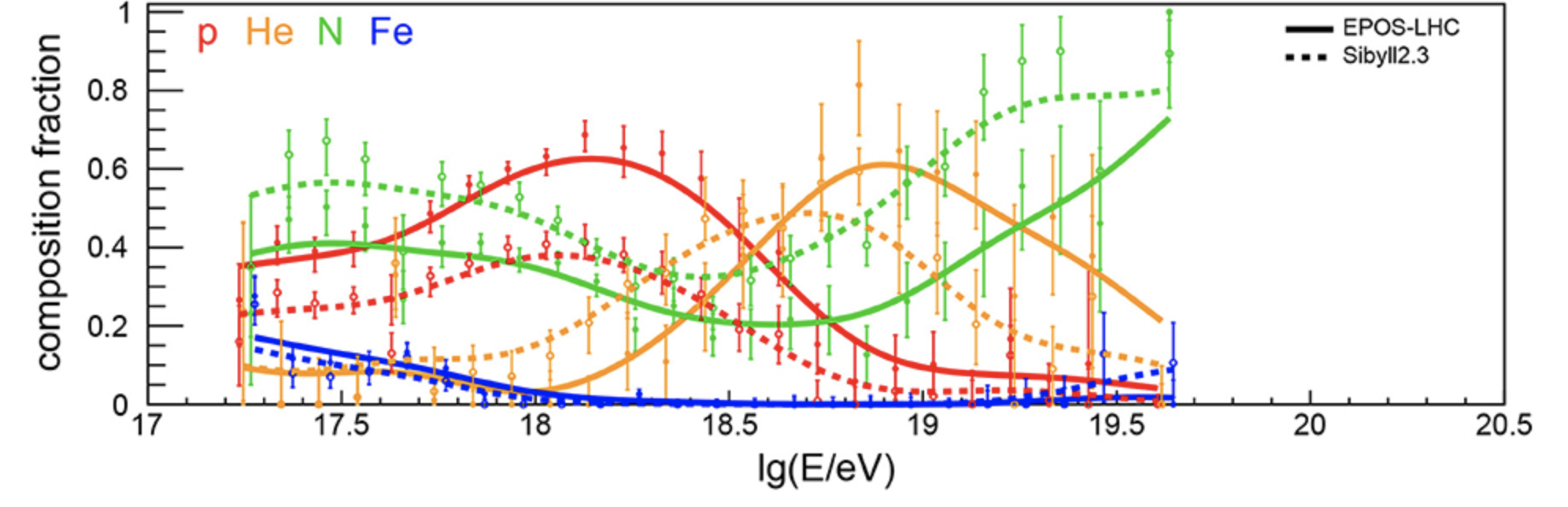}
  \caption{Top: Energy spectrum after combining the individual measurements by the SD-750 and the SD-1500 scaled by $E^{2.6}$ -- from~\cite{PierreAuger:2021hun}. Bottom: Composition fractions observed on Earth derived from fitting templates of four mass groups to the $\Xmax$ distributions -- adapted from~\cite{Bellido:2017cgf}. } 
  \label{fig:spectrum_mass}
\end{figure}

The measurement of the energy spectrum above $100~$PeV is emblematic of the power of using multiple detectors: it has been performed by combining the independent spectra from the two different SD arrays. The different data streams are complementary: data from the SD-750~m array allow for the determination of the energy spectrum down to $100~$PeV~\cite{PierreAuger:2021hun}; data with events from the SD-1500~m array are crucial above the energy of full efficiency of $3~$EeV up to the highest energies~\cite{PierreAuger:2020qqz}. The two spectra, in agreement within uncertainties, are combined into a unique one shown in figure~\ref{fig:spectrum_mass}, top panel, taking into account the systematics of the individual measurements. Beyond the ankle (the hardening at $\simeq 5~$EeV) and suppression (steepening at $\simeq 50~$EeV) spectral features already established previously, the second-knee feature is observed to extend over a wide energy range (not fully covered in this measurement) while a steepening at $\simeq 10~$EeV, dubbed as the instep feature, has been uncovered thanks to the exposure and energy resolution reached at the Observatory. 

The measurement of the depth of the shower maximum, $X_{\mathrm{max}}$, is the most robust mass-sensitive EAS observable. While $X_{\mathrm{max}}$ data cover the region of the ankle well, they do not extend, currently, into the region of flux suppression due to the intrinsically limited duty cycle of the FD. Between $\simeq 150~$PeV and $2~$EeV, $\langle X_{\mathrm{max}}\rangle$ increases by around 77 g~cm$^{-2}$ per decade of energy~\cite{PierreAuger:2019phh}. This is larger than that expected for a constant mass composition (60 g~cm$^{-2}$ per decade) and thus indicates that the mean primary mass is becoming lighter over this energy range. On the other hand, above $2\times10^{18}~$eV, the rate of change of $\langle X_{\mathrm{max}}\rangle$ becomes significantly smaller (26 g~cm$^{-2}$ per decade~\cite{PierreAuger:2019phh}). The trend is thus inverted and the composition gets heavier. The fluctuations of $X_{\mathrm{max}}$ start to decrease above the same energy, $2\times10^{18}~$eV, being rather constant below~\cite{PierreAuger:2019phh}. Composition fractions observed on Earth can be derived from fitting templates of four mass groups to the $\Xmax$ distributions. The results are shown in Fig.~\ref{fig:spectrum_mass} -- adapted from~\cite{Bellido:2017cgf} -- as a function of energy. Error bars denote statistical uncertainties and lines were added to guide the eye. The two interpretations of the data with EPOS-LHC~\cite{Pierog:2013ria} and Sibyll2.3~\cite{riehn2017hadronic} are shown as closed and open symbols with solid and dashed lines styles respectively. Overall, heavier mass groups are observed to take over one after the other so that the all-particle flux gets dominated by one specific mass group in every energy range beyond the ankle energy. As of today, no composition fractions are available around and above $100~$EeV.

\begin{figure}[!t]
  \centering
  \includegraphics[width=0.5\textwidth]{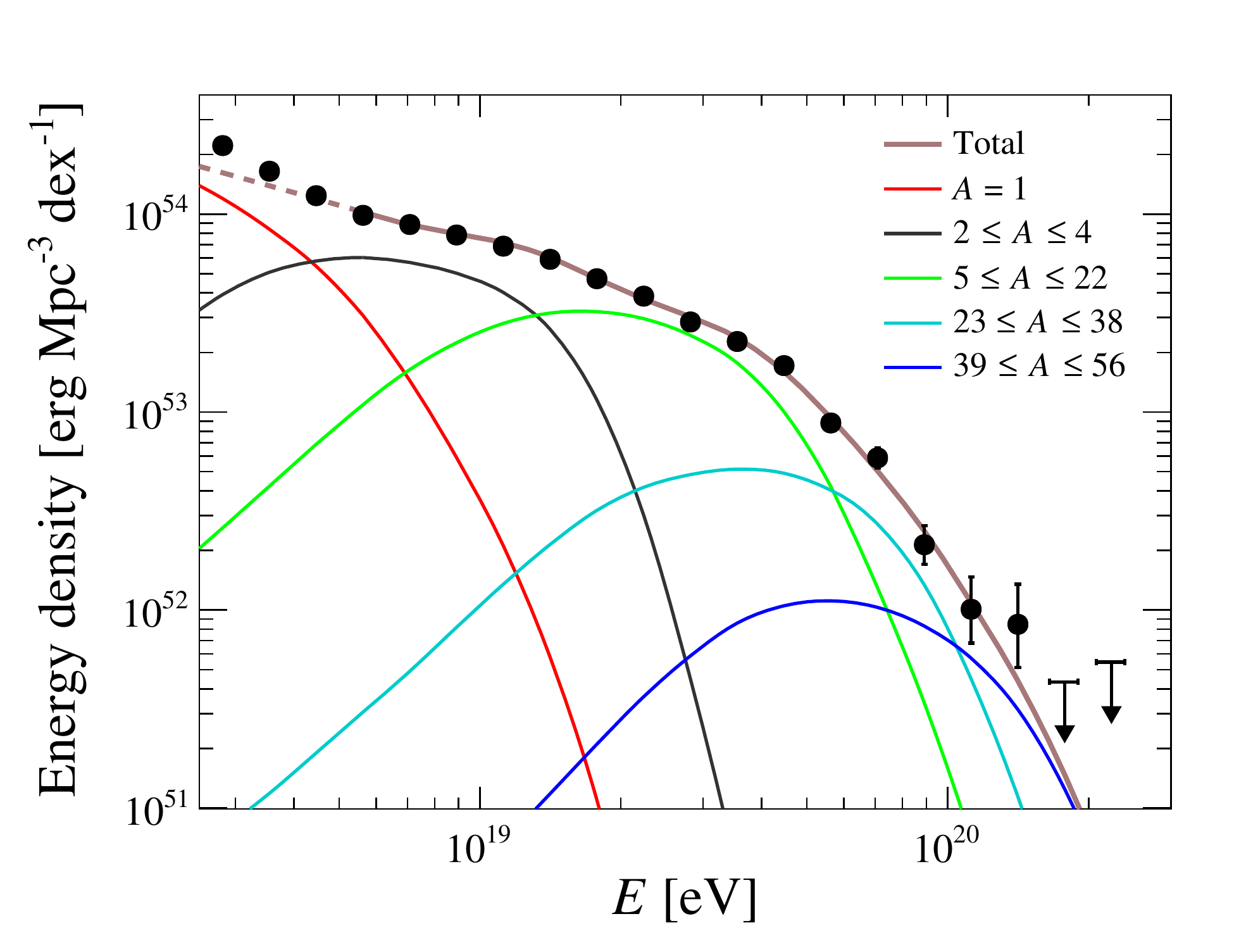}
  \caption{Energy density obtained with the best fit parameters of the benchmark scenario used for illustration, as described in the text. The dashed curve shows the energy range that is not used in the fit and where an additional component is needed for describing the spectrum. From~\cite{PierreAuger:2020kuy}.} 
  \label{fig:e2}
\end{figure}

These results fit a scenario in which several nuclear components contribute to the total intensity and in which  electromagnetic fields permeate source environments where nuclei are accelerated to a maximum energy proportional to their charge ($Z$). Without distraction by the many details a full model scenario would require, we illustrate this scenario in Fig.~\ref{fig:e2} by considering several nuclear components injected at the sources with a power-law spectrum and with the maximal energy of the sources modeled with an exponential cut-off~\cite{PierreAuger:2016use,PierreAuger:2020kuy}. For simplicity, the sources are assumed to be stationary and uniform in a co-moving volume. The best reproduction of the data is shown by simultaneously fitting the energy spectrum above $5~$EeV and the distribution of $X_{\rm max}$ (using EPOS LHC as a  model of hadronic interactions in their interpretation). The abundance of nuclear elements at the sources is dominated by intermediate-mass nuclei accelerated to $\approx 5~Z {\times}~$EeV and escaping from the source environments with a very hard spectral index, which is in turn determined by the quasi-monoelemental increase of the average mass with energy. In this scenario, the steepening observed above $\approx 50$~EeV results from the combination of the maximum energy of acceleration of the heaviest nuclei at the sources and the GZK effect~\cite{Greisen:1966jv,Zatsepin:1966jv}. The steepening at $\approx 10$~EeV reflects the interplay between the flux contributions of the helium and carbon-nitrogen-oxygen components injected at the source with their distinct cut-off energies, shaped by photo-disintegration during the propagation.

\textit{Beyond the UHECR science.} The reach of the data of the Observatory goes beyond the study of UHECRs: measurement of p-air and p-p cross section~\cite{PierreAuger:2012egl}, study of muon number in EAS~\cite{PierreAuger:2016nfk,PierreAuger:2014ucz,PierreAuger:2020gxz,PierreAuger:2021qsd}, study of hadronic interactions~\cite{PierreAuger:2021xah}, searches for multi-messengers of high-energy phenomena in the universe~\cite{PierreAuger:2019fdm,PierreAuger:2020llu,PierreAuger:2021oks}, study of electric phenomena in the atmosphere~\cite{PierreAuger:2020lri,PierreAuger:2021int,PierreAuger:2021ecs}, etc. Among other results, the upper limits obtained on photon~\cite{PierreAuger:2022uwd,PierreAuger:2021mjh,PierreAuger:2022nud} and neutrino~\cite{PierreAuger:2019ens} fluxes in the EeV range and above are of special interest. As an example of the power of  the limits on photon fluxes, we highlight the constraints that can be inferred on the reduced coupling constant of gauge interactions in a dark sector of super-heavy particles. Full details can be found in~\cite{PierreAuger:2022wzk,PierreAuger:2022nud}. Compelling evidence for the observation of the decay of such dark-matter particles would be the detection of a flux of astrophysical photons with energies in excess of ${\simeq}10^8~\mathrm{GeV}$, in particular from regions of denser DM density such as the center of our Galaxy. 

Stability of super-heavy particles is calling for a new quantum number conserved in the dark sector so as to protect the particles from decaying. Nevertheless, even stable particles in the perturbative domain will in general eventually decay due to non-perturbative effects (instantons) in non-abelian gauge theories. Instanton-induced decay can thus 
make observable a dark sector that would otherwise be totally hidden by the conservation of a quantum number~\cite{Kuzmin:1997jua}. Assuming quarks and leptons carry this quantum number and so contribute to anomaly relationships with contributions from the dark sector, they will be secondary products in the decays of SHDM together with the lightest hidden fermion. The lifetime of the decaying particle, $\tau_X \simeq M_X^{-1}\exp{\left(4\pi/\alpha_X\right)}$, is governed by $\alpha_X$, the reduced coupling constant of the hidden gauge interaction~\cite{tHooft:1976rip}. 
As photons would result in the secondaries of the decay process, the limits on their flux can be translated into a bound on the reduced coupling constant: $\alpha_X < 0.09$, for $10^{9} < M_X/{\rm GeV} < 10^{19}$. This upper limit on $\alpha_X$ is the best-ever obtained limit on a proxy of instanton strength. Implications of this result in terms of dark-matter production in the early universe are discussed elsewhere~\cite{PierreAuger:2022nud}. \\

\bibliographystyle{iopart-num}
\bibliography{biblio}
 
\end{document}